# Microwave heating as a universal method to transform confined molecules into armchair graphene nanoribbons

Haoyuan Zhang[1#], Yingzhi Chen[1#], Kunpeng Tang[1#], Ziheng Lin[1], Xuan Li[1], Hongwei Zhang[1], Yifan Zhang[1,2], Takeshi Saito[3], Chi Ho Wong[4], Chi Wah Leung[5], Chee Leung Mak[5], Yuan Hu[6], Weili Cui[1](✉), Kecheng Cao[6](✉), Lei Shi[1](✉)

[1] *State Key Laboratory of Optoelectronic Materials and Technologies, Nanotechnology Research Center, Guangzhou Key Laboratory of Flexible Electronic Materials and Wearable Devices, School of Materials Science and Engineering, Sun Yat-sen University, Guangzhou, 510275, China*
[2] *Huzhou Key Laboratory of Environmental Functional Materials and Pollution Control, School of Engineering, Huzhou University, Huzhou, 313000, China*
[3] *Nanomaterials Research Institute, National Institute of Advanced Industrial Science and Technology (AIST), Tsukuba, Ibaraki 305-8565, Japan*
[4] *Department of Industrial and Systems Engineering, and Research Institute for Advanced Manufacturing, The Hong Kong Polytechnic University, Hong Kong, China*
[5] *Department of Applied Physics, The Hong Kong Polytechnic University, Hong Kong, China*
[6] *Shanghai Key Laboratory of High-resolution Electron Microscopy, ShanghaiTech University, Shanghai, 201210, China*

**ABSTRACT**

Armchair graphene nanoribbons (AGNRs) with sub-nanometer width are potential materials for fabrication of novel nanodevices thanks to their moderate direct band gaps. AGNRs are usually synthesized by polymerizing precursor molecules on substrate surface. However, it is time-consuming and not suitable for large-scale production. AGNRs can also be grown by transforming precursor molecules inside single-walled carbon nanotubes via furnace annealing, but the obtained AGNRs are normally twisted. In this work, microwave heating is applied for transforming precursor molecules into AGNRs. The fast heating process allows synthesizing the AGNRs in seconds. Several different molecules were successfully transformed into AGNRs, suggesting that it is a universal method. More importantly, as demonstrated by Raman spectroscopy, aberration-corrected high-resolution transmission electron microscopy and theoretical calculations, less twisted AGNRs are synthesized by the microwave heating than the furnace annealing. Our results reveal a route for rapid production of AGNRs in large scale, which would benefit future applications in novel AGNRs-based semiconductor devices.

## 1 Introduction

Graphene nanoribbons (GNRs) with a regulable band gap, high mobility, and high carrier capacity are getting increasing attention[1-3]. Especially, armchair GNRs (AGNRs) possess band gaps of 0.1-2.5 eV tuned by the width, making them a potential candidate for the fabrication of novel nanodevices including photodetectors, sensors, and transistors[4-6]. AGNRs can be classified into three groups by the number of dimer lines (n) across their widths: n = 3p, 3p + 1, or 3p + 2, where p is an integer. Importantly, n-AGNRs with n = 3p or 3p + 1 have a comparatively big band gap between 1.2 and 2.3 eV, while nAGNRs with n = 3p + 2 usually have the least band gap in the order of 0.1 eV[7]. Synthesis of n = 3p or 3p + 1 AGNRs attracts attention due to their potentials in semiconductor applications. Especially, n = 6 and 7 AGNRs with widths of 0.62 and 0.74 nm combined with the width-dependent energy gaps of 1.83 and 2.18 eV[8], respectively, are highly concerned. Tailoring both the width and the edge of the AGNRs are keys towards the novel applications.

The GNRs can be prepared by etching graphene into strips or through unzipping carbon nanotubes (CNTs)[9-11]. However, such a top-down approach is not able to achieve GNRs with sub-nanometer in width and controllable edge. In contrast, bottom-up synthesis on surface allows for precise control of both the width and the edge structure of the GNRs by designing precursor molecules as monomers for polymerization[12-14]. However, to obtain specific GNRs, designing and synthesis of monomers are rather complicated and often require multiple steps with low yield. Therefore, on-surface synthesis is time consuming and not suitable for large-scale production.

Confined synthesis using single-walled carbon nanotubes (SWCNTs) with hollow space as nanoreactors enables to synthesize the GNRs. Previously it was proved that polymerizing precursor molecules inside SWCNTs is an effective way to obtain GNRs[15, 16]. Our recent studies indicated that specific AGNRs, e.g., 6-/7-AGNRs, can be formed by filling and transforming the precursor molecules inside SWCNTs [8, 17 18], which became an alternative method of GNR fabrication with precise controls of the width and edge except the on-surface synthesis. We found that the diameter/chirality of the SWCNTs is the key to regulate the structure of the nanoribbons. Most importantly, no specific design is needed for the precursor molecules, since the formation mechanism of the GNRs is completely different from that of the on-surface synthesis. Specifically, in the confined synthesis the molecules were first decomposed into small organic segments, and then reassembled into GNRs under the heating. Thus, the synthesis temperature is usually higher than that of the on-surface synthesis.

Microwave as an energy source enables to heat up the materials instantly. Previously, microwave heating was used for the synthesis,

purification, and functionalization of SWCNTs[19-21]. In this work, microwave heating is utilized to transform confined precursor molecules into 6-/7-AGNRs, which shortens the transformation process, i.e., decomposition and recombination, from hours by furnace annealing into seconds. More importantly, less twisted GNRs are formed by fast heating than the furnace heating, as demonstrated by Raman spectroscopy, aberration-corrected high-resolution transmission electron microscopy (ACHRTEM) and theoretical calculations. Microwave heating provides an effective and efficient way for large-scale production of GNRs with tailored width and edge, which would benefit future applications in novel GNRs-based semiconductor devices.

## 2 Experimental

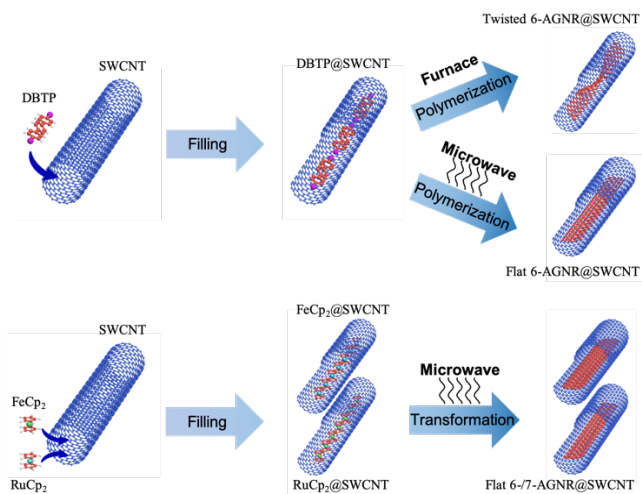

**Figure 1** Illustrations of the transformation from precursor molecules to 6-/7-AGNRs by confined synthesis.

### 2.1 Synthesis of AGNRs

SWCNTs with an average diameter of around 1.3 nm were prepared by enhanced direct injection pyrolysis synthesis (eDIPS)[22]. The as-grown eDIPS SWCNTs were thermally treated in the air at 400 °C for 30 min to remove the amorphous carbon around the catalyst, and then treated with HCl to dissolve the exposed catalysts[23]. SWCNT buckypaper with a thickness of 100 μm was obtained after rinsing with distilled water and drying. The SWCNTs were opened by annealing in air at 500 °C for 30 min. To fill the precursor molecules into the SWCNTs, the opened SWCNTs were sealed with 4,4''-dibromo-$p$-terphenyl (DBTP) or ferrocene ($FeCp_2$) or ruthenocene ($RuCp_2$) in an ampoule under dynamic vacuum of around $2 \times 10^{-3}$ Pa, and heated to their optimal temperatures at 320 or 350 or 250 °C for 3 days, respectively. After filling, the ampoule was placed in a microwave oven (Midea, 700 W, 2450 MHz) and irradiated for tens of seconds to transform the molecules into armchair graphene nanoribbons. The temperature of the sample increases with the irradiation time (Figure S1). For comparison, the molecule-filled samples were annealed in a tube furnace at different temperatures (650-1000 °C) and durations (1-6 hours) under a dynamic vacuum of around $4 \times 10^{-3}$ Pa.

### 2.2 Characterization

The samples were measured by Raman spectroscopy (TriVista 557, Princeton Instruments). The laser wavelengths of 633 and 561 nm were applied to excite the 6-AGNRs and 7-AGNRs, respectively[8]. The laser power was set below 1 mW to prevent the heating effect[17]. All the spectra were calibrated by Rayleigh scattering line and normalized to the intensity of the 2D-band.

The AGNR@SWCNTs samples were characterized by aberration-corrected HRTEM (Thermo Fisher) at the voltage of 60 kV. The exposure time was 1 s for each image.

### 2.3 Simulation methods

The Raman spectra of the samples are implemented by the ab-initio calculations [24]. The infinite 6-AGNR refers to periodic cells. The finite 6-AGNR is obtained from supercell construction. The lateral interaction between the 6-AGNR can be emerged by decreasing the AGNR-to-AGNR distance. The fully relaxed lattice parameters at the spin-restricted PBE level under Projector Augmented Wave pseudopotentials are obtained. The plane-wave energy has a cut-off point at 300 eV[24]. The Raman spectrums are implemented by taking the finite differences for the displacement of the cell per vibrational mode [24].

## 3 Results and Discussion

SWCNTs were proved as effective nanoreactors for confined synthesis of novel 1D carbon materials, including carbon chains[23, 25], carbon nanotubes [26], and GNRs[8, 15, 16] . Especially, 5-/6-/7-AGNRs were successfully synthesized by transforming encapsulated molecules inside SWCNTs via high-temperature annealing in a vacuum by furnace [8, 17, 18]. However, the transformation process usually takes hours and the obtained GNRs are mostly twisted and short. Instead of the furnace, microwave allows for heating up the materials rapidly. Therefore, we applied microwave heating to transform the encapsulated molecules into GNRs. To verify the validity of the method, DBTP molecules were used, since they can be polymerized into 6-AGNRs via surface synthesis at a relatively low temperature of around 460 °C[27, 28]. The DBTP molecules were first reacted into poly-$p$-phenylene, and then GNRs were formed by combining the parallel-arranged poly-$p$-phenylene chains. Therefore, a series of AGNRs with widths of n=6, 9, 12, … can be formed depending on the polymerization temperature[27, 28]. In our case, DBTP molecules were filled inside SWCNTs with an average diameter of 1.3 nm, thus only 6-AGNRs but not 9- and 12-AGNRs can be formed thanks to the confinement of the SWCNTs. Thus, DBTP@SWCNT is an ideal system for testing the confined synthesis.

The DBTP@SWCNTs were annealed in a tube furnace. As shown in Figure 2b, Raman spectra of the annealed sample were taken by a laser with wavelength of 633 nm. Compared with the filled SWCNTs, several Raman modes appear after annealing. Radial breathing-like mode (RBLM) represents the vibrations along the width direction of the AGNRs. The Raman frequency of the RBLM mainly depends on the width of the AGNRs[29]. Here the RBLM locates at around 460 $cm^{-1}$, corresponding to the 6-AGNRs[8]. In addition, the Raman signal at 1244 $cm^{-1}$ is assigned to C-H in-plane bending mode ($CH_{ipb}$). The $CH_{ipb}$ intensity is comparable to that of the G-band, since the laser energy is close to the energy gap of the 6-AGNRs[8], thus strongly resonantly enhanced. Besides, the strong signal suggests the high yield of the 6-AGNRs. The other two Raman modes at 1272 and 1358 $cm^{-1}$ are called defect-like mode (DLM)[8,16], because their frequencies are close to that of the D-band of the SWCNTs. The G-band of the 6-AGNRs is not clearly resolved, because it is weaker than the $CH_{ipb}$ and overlaps with the intense G-band of the SWCNTs. The annealed sample was also measured by a laser with a wavelength of 561 nm, because this laser energy enables the excitation of 7-AGNRs in resonance, which could help to examine the purity of the obtained AGNRs. The width of the 7-AGNRs is only slightly wider than that of the 6-AGNRs, which fits the hollow space of the SWCNTs with a diameter of 1.3 nm as well. However, as shown in Figure 2c, only weak Raman modes of 6-AGNRs can be seen and no signal belonging to the 7-AGNRs is observed, revealing that only 6-AGNRs were synthesized. Then the question is: Can microwave heating enable to polymerize the DBTP into 6-AGNRs.

The DBTP-filled sample from the same branch was annealed via microwave heating. As indicated by the Raman spectra excited with the same two lasers shown in Figure 2, all the new Raman modes of the microwave-annealed sample belong to the 6-AGNRs, which reveals that the microwave heating is truly effective for the polymerization of the DBTP molecules into 6-AGNRs. However, microwave could heat up the sample at a temperature much higher than the optimal polymerization temperature of the DBTP, resulting in a lower yield of the 6-AGNRs. Indeed, the intensities of the RBLM, $CH_{ipb}$ and DLM of the microwave-annealed sample are weaker than those of furnace annealed sample. Furthermore, increasing the microwave annealing duration leads to a decreased yield of the 6-AGNRs and even damages of the GNRs as well as the SWCNTs (Figure S2). All the above results suggest that the growth mechanism (i.e., polymerization) of the 6-AGNRs from DBTP in confined synthesis is the same as the one in the on-surface synthesis, as illustrated in Figure 2a. How about the performance of other precursor molecules treated by the microwave heating?

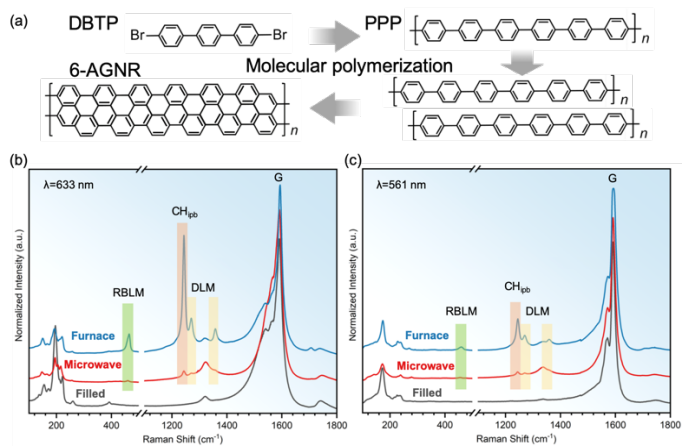

**Figure 2** (a) Polymerization mechanism of DBTP into 6-AGNRs. Raman spectra of 6-AGNRs synthesized from DBTP precursor molecules excited by lasers with wavelengths of (b) 633 nm and (c) 561 nm.

Ferrocene consisting of two cyclopentadienyl rings bound to a central iron atom apparently cannot be polymerized into GNRs with hexatomic rings. However, GNRs were largely formed inside SWCNTs, when the encapsulated ferrocene molecules were annealed in vacuum [16, 17], suggesting that the growth mechanism is different from polymerization. In this case, the ferrocene molecules were first decomposed and then combined into GNRs, thus called "decomposition-recombination" strategy. The width of the synthesized GNRs depends on the diameter of the SWCNTs, but not on the structure and size of the precursor molecule. The confined space inside the SWCNTs facilitates a precise control of the width and edge of the GNRs, giving rise to specific GNRs synthesized, e.g., 6-AGNRs or 7-AGNRs. As shown in Figures 2b and 3b, the $CH_{ipb}$ to G-band ratios of 6-AGNRs obtained from DBTP and ferrocene are similar, meaning that the yield of the 6-AGNRs transformed from ferrocene by furnace annealing is comparable to the 6-AGNRs polymerized from DBTP, which highlights the effectiveness of the "decomposition-recombination" strategy. Interestingly, when the ferrocene is heated by microwave, the yield of the 6-AGNRs (Figure 2b) is much higher than that of the 6-AGNRs polymerized from the DBTP by microwave heating (Figure 3b). This is reasonable, since the optimal transformation temperature (650 °C) of the ferrocene is higher than the polymerization temperature (400 °C) of the DBTP. When the ferrocene is switched into ruthenocene, a similar yield of 6-AGNRs can be observed, as shown in Figure 3d. Except 6-AGNRs, a small quantity of 7-AGNRs from both ferrocene and ruthenocene with furnace annealing or microwave heating were observed, as shown in Figures 3c and 3e, because slightly larger SWCNTs also exist in the SWCNT sample, which are suitable to synthesize the 7-AGNRs.

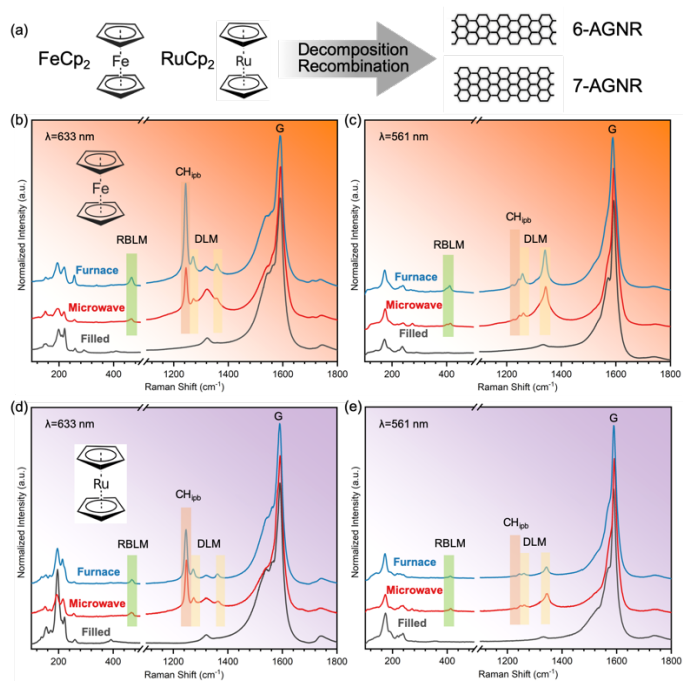

**Figure 3** (a) Schematic diagram of transforming $FeCp_2$ or $RuCp_2$ into 6/7-AGNRs. Raman spectra of 6-AGNRs synthesized from (b) $FeCp_2$ or (d) $RuCp_2$ excited by a laser with wavelength of 633 nm. Raman spectra of 7-AGNRs synthesized from (c) $FeCp_2$ or (e) $RuCp_2$ excited by a laser with wavelength of 561 nm.

To visually verify the formation of 6-/7-AGNRs, the microwave-annealed samples were characterized by ACHRTEM. As shown in Figure 4, two GNRs with the widths of 0.61 and 0.72 nm inside SWCNTs are assigned to 6- and 7-AGNRs, respectively. Note that the shape of the GNRs continuously varied under the electron beam irradiation, because the GNRs freely rotated and translated inside the SWCNTs, which differentiates the edges of the GNRs from the walls of CNTs. The GNRs remained intact with long electron irradiation, suggesting a high stability of the confined GNRs. In general, the GNRs confined inside SWCNTs are mostly twisted[15-17]. In comparison, the GNRs synthesized on the substrate are flat[12-14]. As seen in Figures 4f-4i, the marked area by dashed rectangles clearly recorded the twisted structure, which is consistent with previous results[15]. However, in our observations we found that most of the GNRs in the microwave-annealed sample are flat but not twisted. In addition, the ratio of flat to twisted GNRs in the microwave-annealed sample is higher than the ratio in the furnace-annealed sample. The fast heating in a short time may play a role in forming more flat GNRs. In order to evaluate the enrichment of the flat AGNRs, RBLM should be examined in detail, since the RBLM is very sensitive to the structure of the AGNRs, whereas the $CH_{ipb}$ is more related to the edge of the AGNRs and DLM reflects the internal structure of the AGNRs.

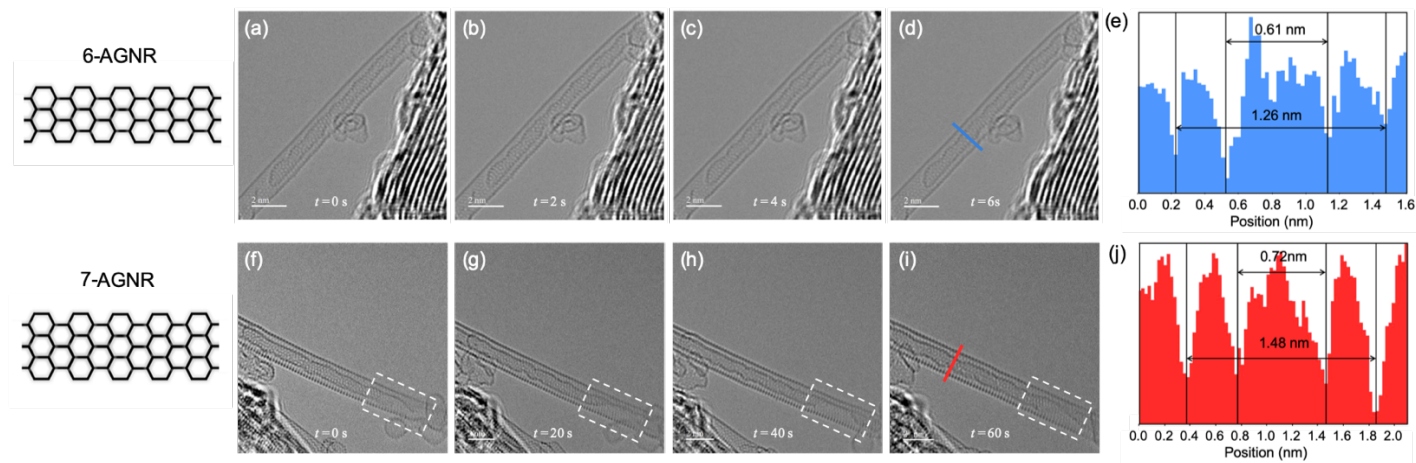

**Figure 4** (a)-(d) ACHRTEM image of selected 6-AGNR@SWCNT. (e) Contrast profile of a 6-AGNR@SWCNT. (f)-(i) ACHRTEM image of selected 7-AGNR@SWCNT. The twisted structure is marked by white dashed rectangles. (j) Contrast profile of a 7-AGNR@SWCNT.

The RBLM frequency mainly depends on the width of the GNRs. The RBLM frequency of ideal AGNRs can be calculated by the equation

$$\omega_{RBLM} = a/w^{1/2} + b$$

where $w$ is the width of the AGNR, and empirical parameters a=1667.9 cm$^{-1}\cdot$Å$^{1/2}$, b=-210.2 cm$^{-1}$ [29]. However, since many of the 6-/7-AGNRs in the sample are not perfect, the RBLM frequency is in principle affected by other factors, e.g., the length of the AGNRs[30], surrounded SWCNTs, and the strain induced from the twist. Indeed, careful observation reveals that the RBLM of the same AGNRs consists of multi-components (Figure 5) and one of the components with lower frequency is consistent with the RBLM frequency of the flat AGNRs obtained from the on-surface synthesis [12, 30], and the other component at higher frequencies could attribute to three factors: the interaction with the SWCNTs, the length of the AGNRs, and/or the twist-induced strain. Since the frequencies of the RBM of the SWCNTs encapsulated with AGNRs did not shift more than 1 cm$^{-1}$ (Figure S3) compared to that of the empty SWCNTs, which is within the spectroscopic resolution, implying that the interaction can be excluded for causing the RBLM splitting of the AGNRs. Similarly, the shifts of the G-band are also small (Figure S4). In addition, theoretical calculations demonstrate that only a small up-shift of the RBLM can be found when considering the interaction, which cannot explain the splitting and big up-shift of the RBLM. Thus, the interaction is not further considered in analyzing the results. As reported in previous work, short AGNRs show a length-dependent Raman signal at around 100 cm$^{-1}$, which was not observed in our samples. Our theoretical calculations indicate that the RBLM of a short 6-AGNR with a length of 1.3 nm splits into several components (Figure 6). All the components shift up or down compared to the RBLM of an infinite 6-AGNR, which is not consistent with our experimental observation. Therefore, we neglect the length effect in our analysis. At last, the third factor should be considered, i.e., the distortion of the GNRs. The distortion of the twisted AGNRs can be certainly expected, as observed in the ACHRTEM, which induces a certain strain in the AGNRs, resulting in up-shifted RBLM, similar to the SWCNTs[31, 32], graphene[33, 34], and AGNRs[35, 36]. Indeed, when we consider 5% strain in the 6-AGNRs, the calculated RBLM splits into two components: One is at a slightly higher position than that of the 6-AGNRs without strain, and the other one locates at a higher frequency. This is completely in line with our experimental observations, as shown in Figure 5.

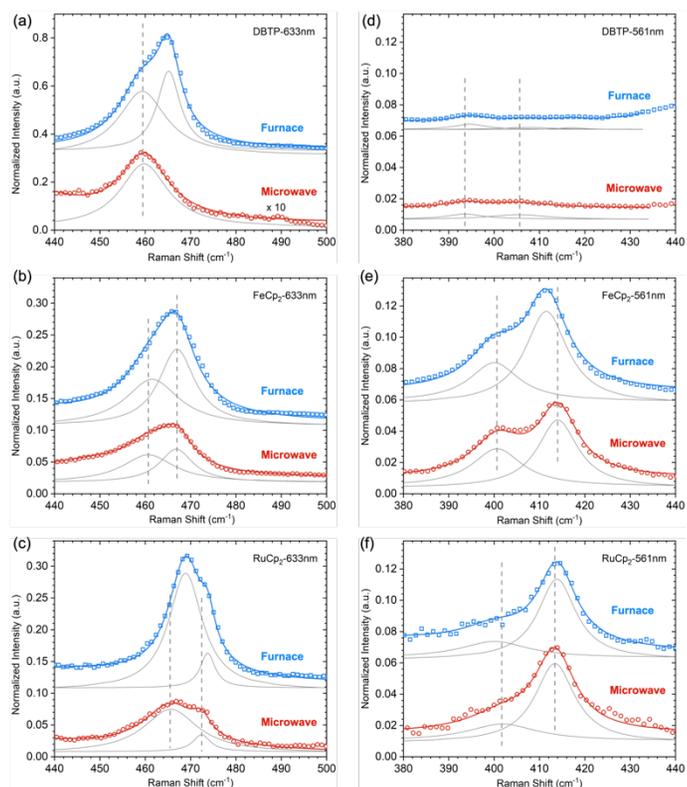

**Figure 5** RBLMs of the transformed 6-AGNRs (left panel) and 7- AGNRs (right panel) from (a, d) DBTP, (b, e) ferrocene, and (c, f) ruthenocene.

With the knowledge of twist-induced RBLM changes, it enables us to analyze the RBLMs of all the samples in detail. As shown in Figure 5, clearly, the RBLM consists of two components for most of the furnace-annealed and microwave-annealed samples. The one at a lower/higher frequency corresponds to flat/twisted AGNRs, which allows to extract the ratio of flat GNRs in a sample by evaluating the area of the component at a lower frequency. The frequencies and ratios of the flat and twisted 6-/7-AGNRs are summarized in Table 1. Comparing the ratios of the flat AGNRs among the samples, we found that microwave annealing for seconds synthesizes a higher ratio of the flat 6-AGNRs than the furnace annealing for hours. Especially, microwave annealing of DBTP@SWCNTs produces almost 100% flat 6-AGNRs, whereas least flat 6-AGNRs can be obtained from the RuCp$_2$@SWCNTs because of higher optimal temperature used. Considering the amount of the frequency shift in the experiment and the calculated results, we evaluate that around 1-2% stain exists in the twisted 6-AGNRs.

In order to check how the furnace annealing plays a role in transforming flat 6-AGNRs into twisted 6-AGNRs, temperature-dependent and time-dependent synthesis of 6-AGNRs from the DBTP@SWCNTs and RuCp2@SWCNTs were performed. As shown in Figure 7 and Figure S5, increasing the annealing temperature and duration lead to transforming the flat 6-AGNRs into twisted 6-AGNRs, which suggests that the twisted 6-AGNRs are more stable inside the SWCNTs when compared to the flat 6-AGNRs. Furthermore, annealing temperature higher than 900 °C ultimately damages the 6-AGNRs. Therefore, fast annealing via microwave is superior for obtaining flat GNRs than the furnace annealing, which could also overcome the large-scale synthesis that encountered in the on-surface synthesis.

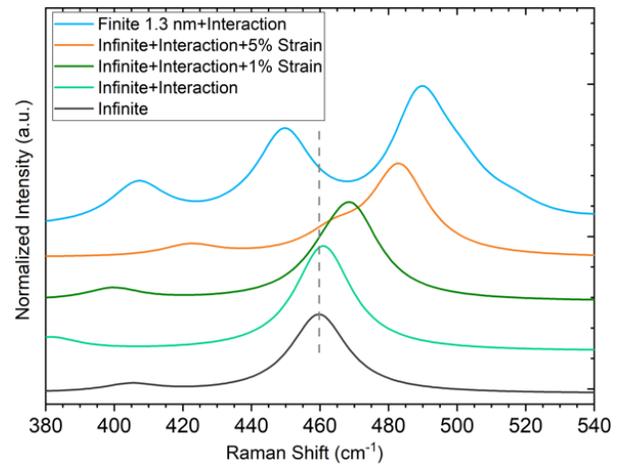

**Figure 6** Theoretical calculations of the RBLMs of the infinite 6-AGNRs with interaction/strain and finite 6-AGNRs with a length of 1.3 nm. The 'infinite' refers to an isolated infinite 6-AGNR, where the lateral spacing is above ~1 nm. The interaction between the infinite 6-AGNRs is emerged in the simulation by setting the lateral interaction to ~0.5 nm, in which the laterally interacted sample is marked as 'infinite+interaction'. The laterally interacted samples are strained by 1% and 5% along the longitudinal axis, respectively. The finite 6-AGNR in the length of 1.3 nm is laterally spaced by ~0.5 nm. and the sample is named in the short form of 'finite 1.3 nm+interaction'. The dashed line works as a guide to the eye.

**Table 1** RBLM frequencies of 6-AGNRs and 7- AGNRs. The unit for the Raman shift is cm$^{-1}$. Ratios of the flat AGNRs in the samples are listed in parenthesis.

| AGNRs | Theory | On-surface synthesis | Furnace annealing | | | Microwave heating | | |
|---|---|---|---|---|---|---|---|---|
| | | | DBTP (ratio) | FeCp$_2$ (ratio) | RuCp$_2$ (ratio) | DBTP (ratio) | FeCp$_2$ (ratio) | RuCp$_2$ (ratio) |
| **6-AGNRs** | 466[29] 460 (this work) | No record yet | 459/465 (62%) | 461/467 (48%±20%) | 469/474 (88%±5%) | 459 (100%) | 461/467 (57%±14%) | 465/472 (89%±9%) |
| **7-AGNRs** | 403[29] | 399[37] | N/A | 400/412 (33%±1%) | 401/414 (26%±6%) | N/A | 401/414 (40%±2%) | 402/413 (27%±6%) |

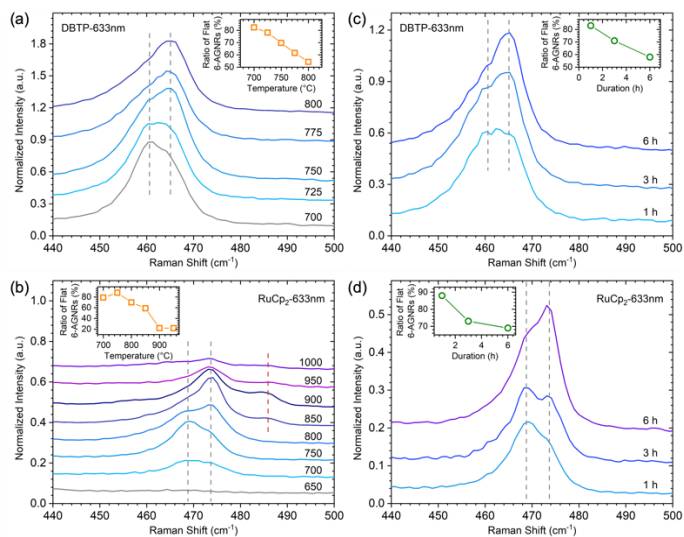

**Figure 7** RBLMs of the 6-AGNRs transformed from (a) DBTP and (b) ruthenocene at different temperatures, from (c) DBTP and (d) ruthenocene with different durations.

## Conclusion

In conclusion, we found that microwave heating, instead of furnace annealing, can be utilized to not only polymerize the DBTP molecules as monomer into 6-AGNRs but also transform the ferrocene and ruthenocene as precursor molecules into 6-AGNRs through "decomposition-recombination" strategy. Even better, the microwave heating allows to obtain a higher ratio of flat 6-AGNRs in seconds than the furnace annealing in hours, as demonstrated by ACHRTEM observations and Raman spectroscopy. Furthermore, microwave heating enables the macroscopic preparation of the 6-AGNRs, which would benefit future applications using the 6-AGNRs as semiconductors with moderate energy gaps.

## Acknowledgements

This work was supported by Guangzhou Basic and Applied Basic Research Foundation (202201011790), Guangdong Basic and Applied Basic Research Foundation (2019A1515011227), National Natural Science Foundation of China (51902353), Fundamental Research Funds for the Central Universities, Sun Yat-sen University (22lgqb03) and State Key Laboratory of Optoelectronic Materials and Technologies (OEMT-2022-ZRC-01). We thank the Department of Applied Physics at The Hong Kong Polytechnic University to provide the ab-initio supports. We also thank the Research Institute for Advanced Manufacturing at The Hong Kong Polytechnic University.

**Electronic Supplementary Material**: The data that support the findings of this study are available from the corresponding author upon reasonable request.
http://dx.doi.org/10.1007/s12274-***-****-* (automatically inserted by the publisher).

# Electronic Supplementary Material

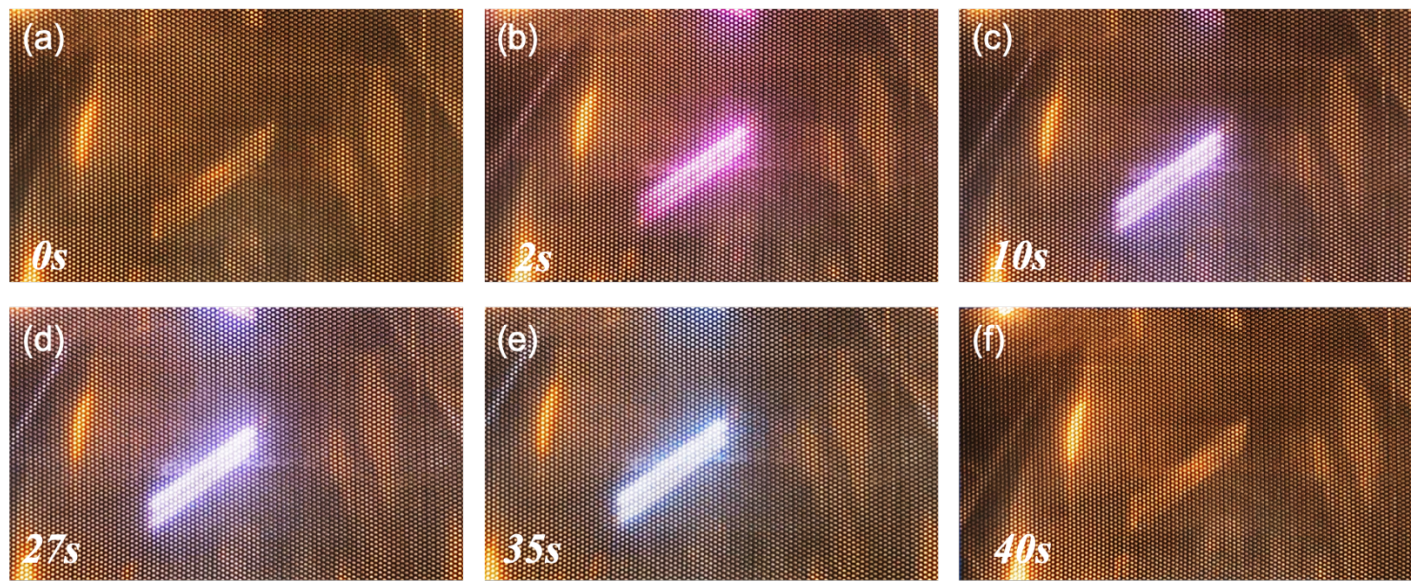

**Figure S1** Optical images of the ruthenocene-filled SWCNTs treated with microwave at different time. The color of the emitting light changes due to the increased temperature.

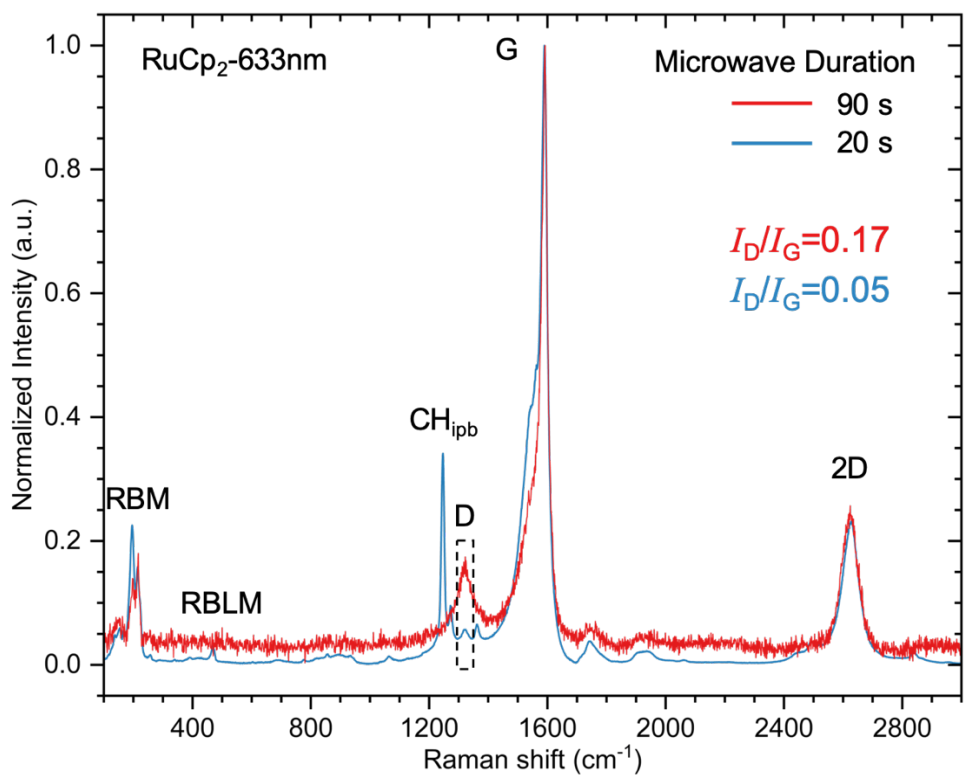

**Figure S2** Raman spectra of the ruthenocene-filled SWCNTs heated by microwave with different durations excited by a laser with wavelength of 633 nm.

Address correspondence to Weili Cui, cuiwli@mail.sysu.edu.cn; Kecheng Cao, caokch@shanghaitech.edu.cn; Lei Shi, shilei26@mail.sysu.edu.cn

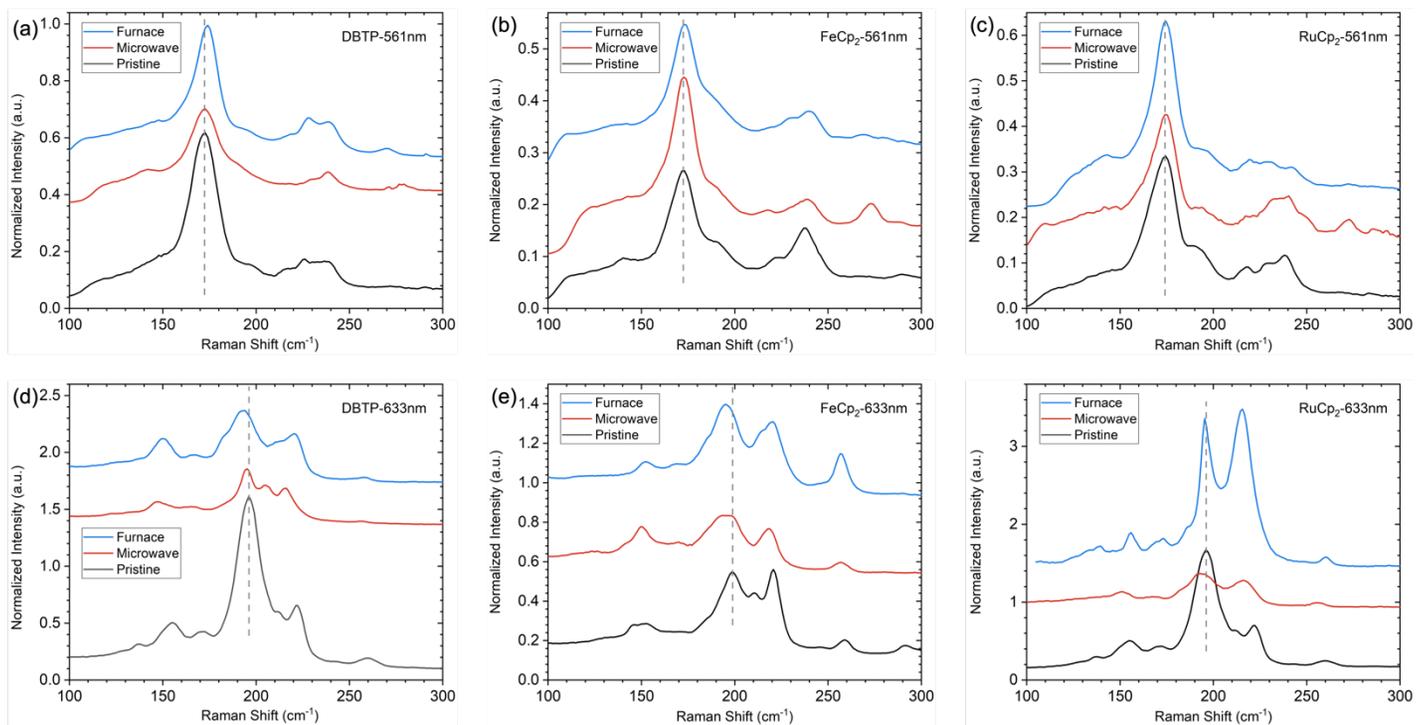

**Figure S3** Raman spectra in the RBM region of the pristine (empty), microwave-annealed and furnace-annealed SWCNT samples excited by two lasers with wavelengths of 561 and 633 nm. (a,d) DBTP, (b,e) FeCp$_2$ and (c,f) RuCp$_2$.

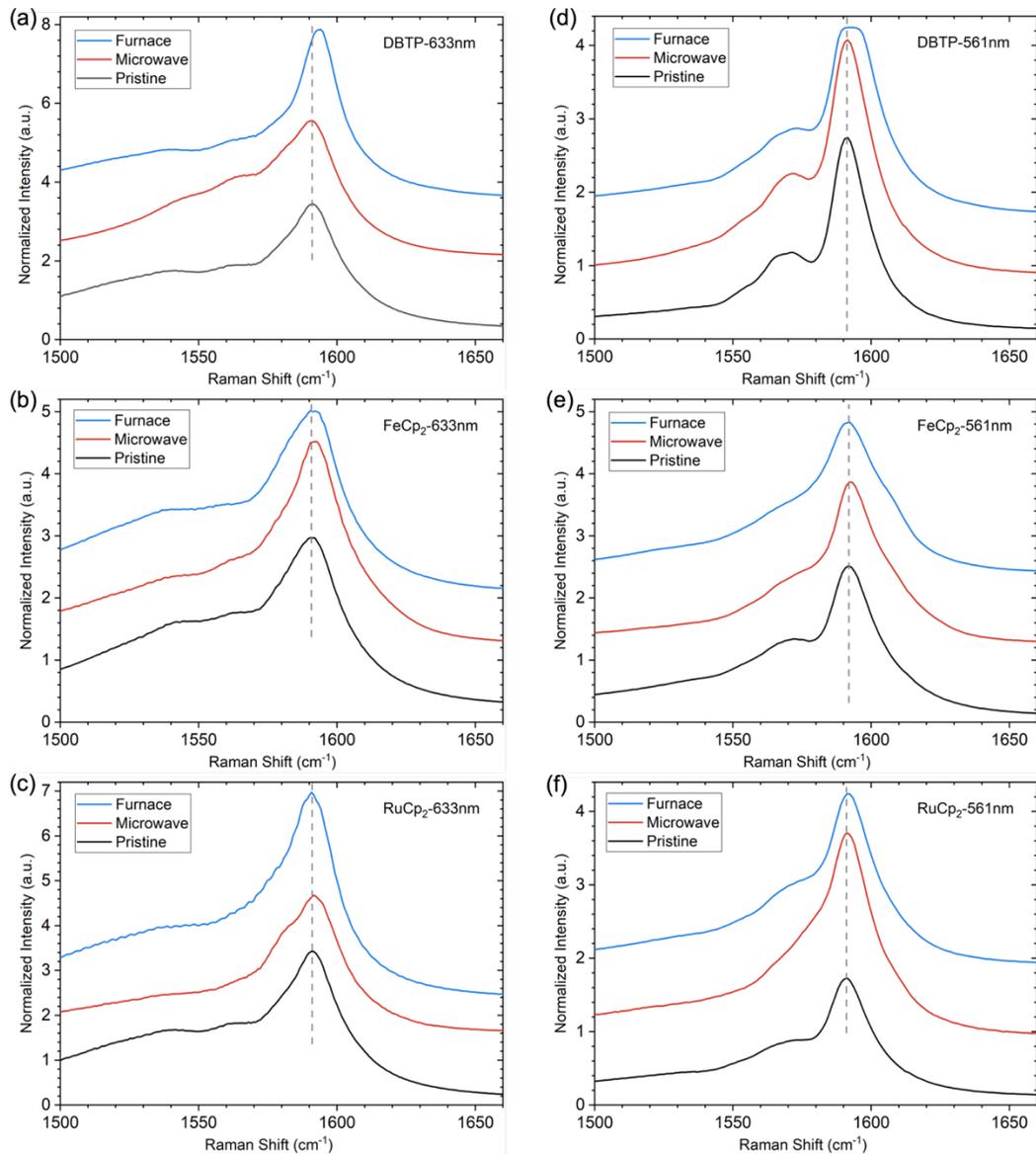

**Figure S4** Raman spectra in the G-band region of the empty, microwave-annealed and furnace-annealed SWCNT samples from (a, d) DBTP, (b, e) ferrocene, and (c, f) ruthenocene excited by two lasers with wavelengths of 561 and 633 nm.

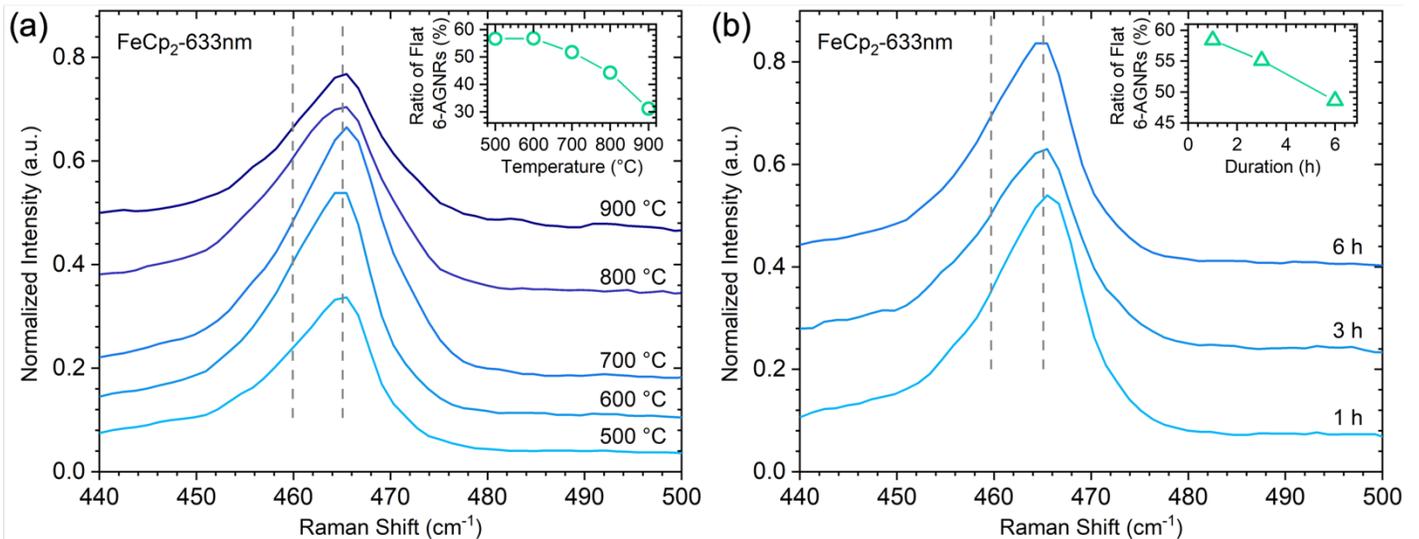

**Figure S5** RBLMs of the 6-AGNRs transformed from ferrocene (a) at different temperatures and (b) with different durations excited by a laser with wavelength of 633 nm.